\documentclass[sigconf]{acmart}
\usepackage{booktabs} 
\usepackage{array}
\usepackage{placeins}

\usepackage{url}  
\usepackage{graphicx}  
\usepackage{mathtools}
\usepackage{amssymb}

\usepackage{amsmath}
\usepackage{upgreek}
\usepackage{multirow}
\usepackage{adjustbox}
\usepackage{pdfpages}

\setcopyright{none}

\begin{document}
\title[Sequential Skip Prediction via RNNs]{Session-based Sequential Skip Prediction via Recurrent Neural Networks}

\author{Lin Zhu}
\affiliation{%
  \institution{Ctrip Travel Network Technology Co., Limited.}
  \city{Shanghai} 
  \state{P.R. China} 
}
\email{lizhonyx@163.com}

\author{Yihong Chen}
\affiliation{%
  \institution{Ctrip Travel Network Technology Co., Limited.}
  \city{Shanghai} 
  \state{P.R. China} 
}
\email{yihongchen@Ctrip.com}

\renewcommand{\shortauthors}{L. Zhu et al.}

\copyrightyear{2019} 
\acmYear{2019} 
\setcopyright{acmcopyright}
\acmConference[WSDM Cup '19]{Proceedings of the ACM WSDM Cup 2019}{Feb 11, 2019}{Melbourne, Australia}

\begin{abstract}
	
The focus of WSDM cup 2019 is session-based sequential skip prediction, i.e. predicting whether users will skip tracks, given their immediately preceding interactions in their listening session. This paper provides the solution of our team \textbf{ekffar} to this challenge. We focus on recurrent-neural-network-based deep learning approaches which have previously been shown to perform well on session-based recommendation problems. We show that by choosing an appropriate recurrent architecture that properly accounts for the given information such as user interaction features and song metadata, a single neural network could achieve a Mean Average Accuracy (AA) score of 0.648 on the withheld test data. Meanwhile, by ensembling several variants of the core model, the overall recommendation accuracy can be improved even further.
By using the proposed approach, our team was able to attain the 1st place in the competition. We have open-sourced our implementation at GitHub\footnote{https://github.com/linzhu123455/spotify-skip-prediction-top-1-solution}.
\end{abstract}

%
%




\maketitle

\section{Introduction}


The accurate prediction/modeling of user action has long been a topic of vital importance in large-scale web and mobile applications, the quality of the model or accuracy of prediction plays a vital role in various revenue generating applications, such as content recommendation \cite{covington2016deep}, advertisement displaying \cite{juan2017field}, search result ranking \cite{mcmahan2013ad}, etc. Traditional approaches, such as matrix Factorization \cite{koren2009matrix} and its variants, seek to model ‘global’ interactions, e.g. by learning low-dimensional user and item embeddings and predicting their interactions in the embedding space. While these algorithms are able to effectively model static user preferences, they fail to account for sequential and dynamical nature of user behavior. This limitation motivates the study of \textit{session-based} user action modeling problem, where the terminology `session' refers to a group of interactions that take place within a given time interval. A session usually has a goal, such as finding a good restaurant in a city, or listening to music of a certain style or mood.

In this year, the WSDM Cup, a competition held annually as part of the prestigious ACM International Conference on Web Search and Data Mining (WSDM), challenges competitors from all over the world to help and build a better user behavior modeling system \cite{brost2019music}. The challenge dataset which consists of roughly 130 million music listening sessions with associated user interactions, was kindly produced by Spotify, a major online music streaming service with online music streaming service with over 190 million active users interacting with a library of over 40 million tracks. The task is to predict whether individual tracks encountered in the second half of a listening session will be skipped by a particular user. 

To tackle this challenge, we focus on recurrent-neural-network-based deep learning approaches since they are able to deal with large datasets in an efficient and effective way, and have previously been shown to perform well on session-based recommendation problems. We show that by choosing an appropriate recurrent architecture that properly accounts for the given information such as user interaction features and song metadata, a single neural network could achieve a Mean Average Accuracy (AA) score of 0.648 on the withheld test data, achieving top 1 position on the challenge leaderboard. Meanwhile, by ensembling several variants of the core model, the overall recommendation accuracy can be improved even further. 

The paper is organized as follows. A overview of the WSDM-Spotify challenge is presented in Section 2, Section 3 discusses the components of our approach, and the paper concludes in Section 4.

\section{The Challenge Description}

\subsection{Dataset}
The public part of the dataset consists of roughly 130 million listening sessions with associated user interactions and context features on the Spotify service. The user interaction with each listened song is described by a set of features, such as the skip behavior, as well as the number of times the user did a seek forward or seek back within track (``hist\_user\_behavior\_n\_seekfwd'' and ``hist\_user\_behavior\_n\_seekback'', respectively), on the other hand, the context features include the hour of day that the listening happened (``hour\_of\_day'') and the type of context the playback occurred within
(``context\_type''), etc. In addition to the public part of the dataset, approximately 30 million listening sessions are used for the challenge leaderboard. For these leaderboard sessions the participant is provided all the user interaction features for the first half of the session, but only the track id’s for the second half. Meanwhile, the meta information of the tracks that users interacted with during these sessions are also provided, including numerous acoustic features and metadata (such as duration and release year).

\subsection{Challenge}
The task is to predict whether individual tracks encountered in a listening session will be skipped by a particular user. In order to do this, complete information about the first half of a user’s listening session is provided, while the prediction is to be carried out on the second half. Participants have access to metadata, as well as acoustic descriptors, for all the tracks encountered in listening sessions. The output of a prediction is a binary variable for each track in the second half of the session indicating if it was skipped or not. 

\subsection{Evaluation Metric}

Although it would also be useful to predict the skip behavior across the entire session, as skip prediction can be used to avoid recommendation of a potential track to the user based on the user's immediately preceding interactions, accurate prediction of whether the next immediate track is going to be skipped is therefore most important. Based on these considerations, Mean Average Accuracy is adopted as the primary evaluation metric for the challenge, with the average accuracy defined by
\begin{equation}
	AA=\frac{\sum\limits_{i=1}^{T}{A\left( i \right)L\left( i \right)}}{T},
\end{equation}
where \textit{T} is the number of tracks to be predicted for the given session, $A\left( i \right)$ is the accuracy at position \textit{i} of the sequence, $L\left( i \right)$ is the boolean indicator for if the \textit{i}th prediction was correct.

\section{Our Solution}

\subsection{Problem Formulation}

The given information about the first half of a session be summarized as a series of triplets ${{\left\{ \left( {{S}_{i}},{{A}_{i}},{{R}_{i}} \right) \right\}}_{1\le i\le 10}}$, where ${{S}_{i}}$, ${{A}_{i}}$, and ${{R}_{i}}$ denote the song, the user interaction associated with the song, and the position of song listening event within the session, on the other hand, information provided about the second half of a session can be summarized as a series of doublets ${{\left\{ \left( {{S}_{i}},{{R}_{i}} \right) \right\}}_{11\le i\le 20}}$, with the user interaction information held out. The objective of the competition is to predict the user skip behavior associated with songs in the second half of the session. 
To integrate the provided information in a principled way, we parametrize the desired prediction model as a neural
architecture that consists of three main components, as illustrated in Figure 1:
\begin{figure}
	\centering
	\includegraphics[width=8cm]{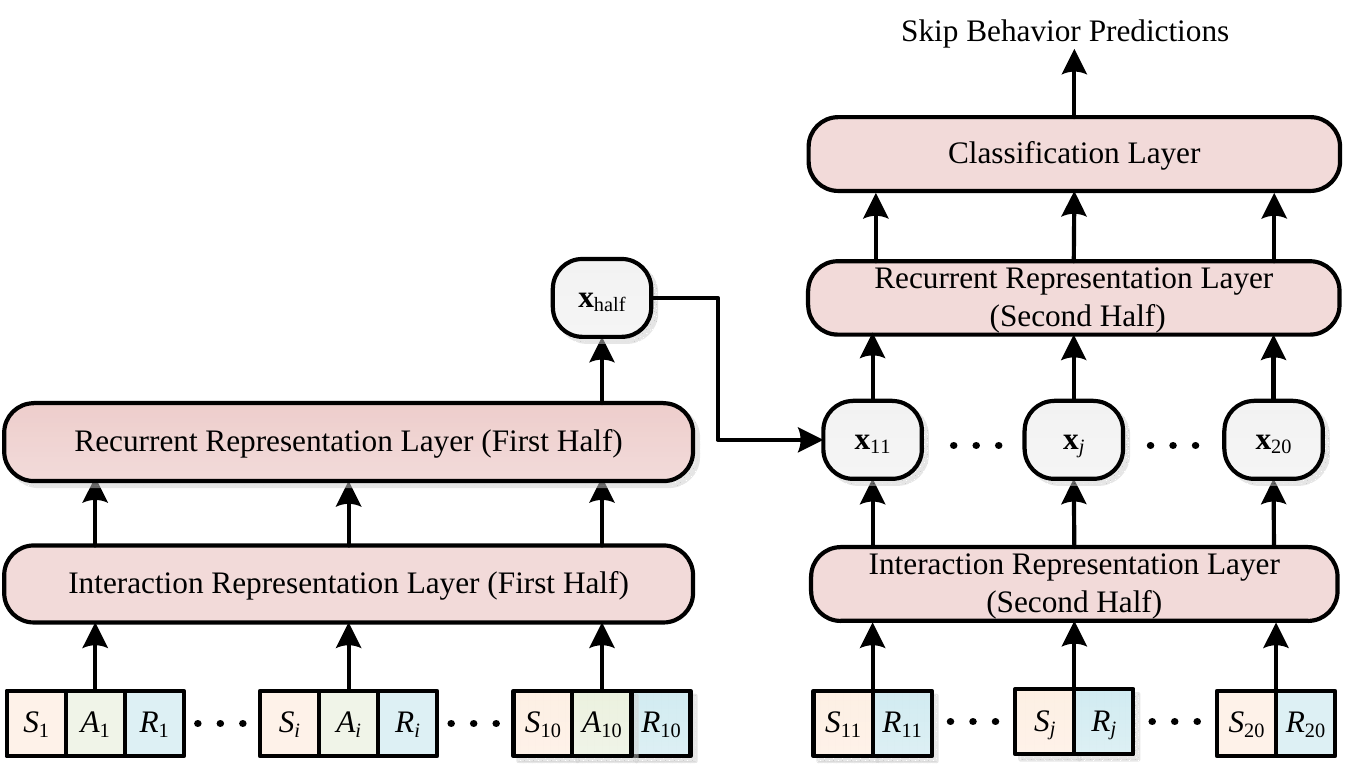} 
	\caption{The overall architecture of our skip prediction model.}
	\label{Architecture}
\end{figure}

\begin{itemize}
	\item \textbf{Interaction Representation Layers} that transforms triplets in the first half of a session and doublets in the second half of a session into feature representations;
	\item \textbf{Recurrent Representation Layers} that utilize recurrent neural network models to refine the triplet and doublet representations by exploiting the sequential nature of user sessions;
	\item A \textbf{Classification Layer} that aggregate the outputs of previous layers and calculate skip probabilities for songs in the second half of the session. 
\end{itemize}

\subsection{Interaction Representation Layers}

As mentioned in Section 2, features for user interactions and the tracks that users interacted with during these sessions are provided, which can be straightforwardly used as their representations. Given the sensitivity of neural networks to input scaling \cite{lecun2012efficient,covington2016deep}, numerical features are normalized to the interval $\left[ 0,1 \right]$, and categorical features such as `context\_type' are mapped to dense vectors via embeddings, which are learned jointly with all other model parameters through back-propagation. 

Additionally, we borrowed ideas from \cite{grbovic2015commerce,grbovic2018real}, and treated each track as a word and each listening session as a sentence, Glove \cite{pennington2014glove} was then used to train on these ``sentences'' to obtain additional 150-dimensional embeddings for songs.

As for the session position information, we simply incorporate it as an additional numerical feature. While more sophisticated positional encoding schemes \cite{vaswani2017attention} are readily applicable, we found that such a simple option already performs well on the challenge dataset.

All of the aforementioned features are accordingly concatenated to obtain vector representations for triplets ${{\left\{ \left( {{S}_{i}},{{A}_{i}},{{R}_{i}} \right) \right\}}_{1\le i\le 10}}$ in the first half of the session, as well as doublets ${{\left\{ \left( {{S}_{i}},{{R}_{i}} \right) \right\}}_{11\le i\le 20}}$ in the second half of the session.

\subsection{Recurrent Neural Networks for Session Modeling}

Motivated by the tremendous success of deep neural networks (DNN) in a number of tasks
such as image and natural language processing (NLP), various approaches have been proposed in recent years to adapt DNN algorithms for data mining tasks \cite{he2017neural,cheng2016wide,covington2016deep}. In particular, session-based user action modeling problem shares some similarities with NLP-related problems \cite{hidasi2015session} as they both deals with sequences. As a result, the Recurrent Neural Networks (RNNs), which have proven to be suitable for handling various sequence modeling tasks \cite{graves2013speech,mikolov2010recurrent,sutskever2014sequence,ha2017neural}, have also been successfully applied to session-based user action modeling, achieving remarkable results of outperforming traditional recommendation methods by 15\% to 30\% in terms of various evaluation metrics \cite{manotumruksa2018contextual,hidasi2015session,LatentCross,donkers2017sequential, quadrana2017personalizing}.

The main difference between RNNs and conventional feedforward neural network models is the existence of an internal hidden state in the units that recurrently process and memorize sequences of inputs. Traditional RNN models usually suffer from the vanishing gradient problem when the models need to process long sequences. The Gated Recurrent Unit (GRU) network is a technique proposed by Cho et al. \cite{chung2014empirical} which aims to solve the problem of gradient vanishing in RNN. Its basic idea is to control the update of network states with an update gate and a reset gate. Formally, let ${{\mathbf{x}}_{t}}\in {{\mathbb{R}}^{\alpha }}$ be the input vector in the \textit{t} step and $\alpha$ is the dimensionality of ${{\mathbf{x}}_{t}}$, The output vector (also the activation vector in GRU) ${{\mathbf{o}}_{t}}\in {{\mathbb{R}}^{\alpha }}$ and the network state ${{\mathbf{s}}_{t}}\in {{\mathbb{R}}^{\alpha }}$ is computed as: 

\begin{equation}
	\begin{matrix}
	{{\mathbf{o}}_{t}}=\left( 1-{{\mathbf{u}}_{t}} \right)\odot {{\mathbf{o}}_{t-1}}+{{\mathbf{u}}_{t}}\odot {{\mathbf{s}}_{t}}, \\ 
	{{\mathbf{u}}_{t}}=\sigma \left( \mathbf{W}_{u}^{x}\cdot {{\mathbf{x}}_{t}}+\mathbf{W}_{u}^{s}\cdot {{\mathbf{s}}_{t}} \right), \\ 
	{{\mathbf{s}}_{t}}=\tanh \left( {{\mathbf{W}}^{x}}\cdot {{\mathbf{x}}_{t}}+{{\mathbf{W}}^{s}}\cdot \left( {{\mathbf{r}}_{t}}\odot {{\mathbf{o}}_{t-1}} \right) \right), \\ 
	{{\mathbf{r}}_{t}}=\sigma \left( \mathbf{W}_{r}^{x}\cdot {{\mathbf{x}}_{t}}+\mathbf{W}_{r}^{s}\cdot {{\mathbf{s}}_{t}} \right), \\ 
	\end{matrix}
\end{equation}

Based on the above considerations, we adopted GRUs to model user sessions. Concretely, for the first half of the  session, we used triplet representations described in the previous section as input, and passed them through 2 stacked GRU networks, the final states of which were then concatenated as a vector ${{\mathbf{x}}_{\text{half}}}$ to represent the entire first half session. Meanwhile, for the second half of the session, let the doublet representation for session position \textit{i} be denoted as ${{\mathbf{x}}_{i}}$, we enrich it with  ${{\mathbf{x}}_{\text{half}}}$ as:
\begin{equation}
\label{enrich}
{{\mathbf{x}}_{i}}=\left[ \begin{matrix}
{{\mathbf{x}}_{i}}  \\
{{\mathbf{x}}_{\text{half}}}  \\
{{\mathbf{x}}_{\text{half}}}\odot \text{ReLU}\left( {{\mathbf{x}}_{i}} \right)  \\
\end{matrix} \right],
\end{equation}
where $\odot$ is the Hadamard product, ReLU is a dense layer with Rectified Linear Units (ReLU) activations \cite{glorot2011deep} to match the dimensions of ${\mathbf{x}}_{i}$ and ${\mathbf{x}}_{\text{half}}$. ${{\mathbf{x}}_{i}}$ returned by (\ref{enrich}) can be considered as an enriched representation as it encodes the contextual information from the first and the second half of the session, as well as the information associated with the specific session position \textit{i}, and all of these information can be beneficial for the considered skip prediction task.

\subsection{Classification Layer}

Given the enriched representation (\ref{enrich}) as input, the classification layer is simply a multilayer perception with sigmoid output activation function and 2 layers of ReLU activations. Meanwhile, to inject more information into the model, we further adopt a multi task scenario where model not only predicts the user skip behavior, but also other user interactions such as 'context\_switch', 'no\_pause\_before\_play', and  'short\_pause\_before\_play'.

\subsection{Implementation Details}

The neural network model is implemented using Keras, a diagram of the model drawn using the Keras function "plot\_model" is given in Fig.2. Note that the user sessions can have unequal length, varying between 10 and 20 in the challenge dataset, we resolve this issue by padding short sequences with pre-specified constants. The model was trained with the Adam algorithm \cite{kingma2014adam} with an initial learning rate of 0.0005, batch size of 2200. 

\begin{figure*}
	\centering
	\includegraphics[width=14cm]{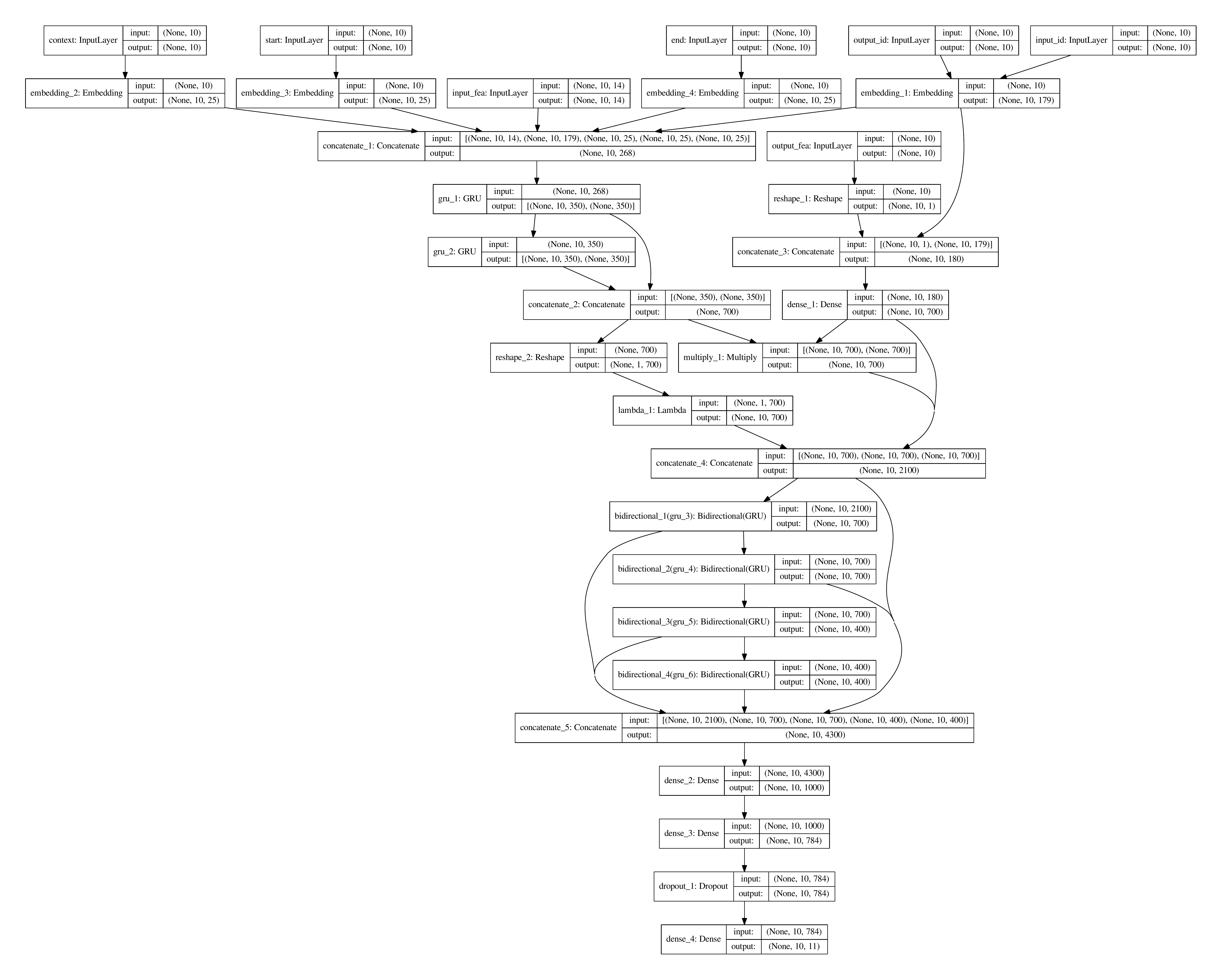} 
	\caption{Model diagram.}
	\label{modelplot}
\end{figure*}

\subsection{Results}

Based on the neural network architecture, a single trained model could achieve a AA score of 0.648 on the withheld test dataset, already achieving top 1 position on the competition leaderboard, we also tried several simple variants of the core model, for example by replacing the ReLU activations with exponential linear units (ELUs) \cite{elu}, or by adding batch normalization layer \cite{ioffe2015batch} before the activations, or by varying batch size and number of hidden units. An ensemble of 6 such model variants could further boost the leaderboard accuracy to 0.651.

\section{Conclusion}

In this paper, we describe our solution to WSDM cup 2019, we focus on recurrent-neural-network-based deep learning approaches since they have previously been shown to perform well on session-based recommendation problems. We show that by choosing an appropriate recurrent architecture that properly accounts for the given information such as user interaction features and song metadata, a single neural network could achieve top 1 position on the challenge leaderboard. Meanwhile, by ensembling several variants of the core model, the overall recommendation accuracy can be improved even further.

\bibliographystyle{ACM-Reference-Format}
\bibliography{sigproc} 

\end{document}